\documentclass[aps,prl,final,twocolumn,showpacs,superscriptaddress]{revtex4-1}
\usepackage{graphicx}
\begin{document}
\title{Anomalous magnetic anisotropy of the topmost surface layer of Ni(110)}
\author{Marco Busch}
\email[Corresponding author: ]{mbusch@physik.hu-berlin.de}
\affiliation{Institut f\"ur Physik, Humboldt-Universit\"at zu Berlin, Newtonstrasse 15, D-12489 Berlin, Germany.}
\author{Jens Lienemann}
\affiliation{Institut f\"ur Physik, Humboldt-Universit\"at zu Berlin, Newtonstrasse 15, D-12489 Berlin, Germany.}
\author{Michael Potthoff}
\affiliation{I. Institut f\"ur Theoretische Physik, Universit\"at Hamburg, Jungiusstrasse 9, D-20355 Hamburg, Germany.}
\author{Helmut Winter}
\affiliation{Institut f\"ur Physik, Humboldt-Universit\"at zu Berlin, Newtonstrasse 15, D-12489 Berlin, Germany.}
\date{\today}
\begin{abstract}
The orientation of the magnetization of a Ni(110) surface was investigated using techniques with different probing depths. By making use of electron capture into excited states of fast He atoms, we found that the magnetization of the topmost surface layer is not aligned along the easy axes of Ni.  However, for a 50 ML film Fe on Ni(110) we observed the magnetization of the topmost Fe surface layer is along the easy axes of Fe.
\end{abstract}
\maketitle
The magnetic anisotropy is the dependence of the magnetic anisotropy energy (MAE) on direction of spontaneous magnetization. In general, magnetic anisotropy is related to the symmetry of a crystal known as magnetocrystalline anisotropy. In the absence of an external magnetic field, a magnetically anisotropic material will align its total moment along an {\it easy axis}, which is an energetically favorable direction of the spontaneous magnetization. The two opposite directions along an easy axis are usually equivalent, and the actual direction of magnetization can be either of them. The orientations of the easy and hard axes of the elemental 3d-Ferromagnets Fe, Co and Ni were experimentally well established \cite{Kaya1,Kaya2,Kaya3}. The easy axes are $<$100$>$ for Fe, $<$0001$>$ for Co and $<$111$>$ for Ni. They are characterized by a small magnitude of an external magnetic field in order to achieve saturation magnetization. In the case of Ni(110) both easy axes $[111]$ or $[\bar{1}\bar{1}\bar{1}]$ and $[11\bar{1}]$ or $[\bar{1}\bar{1}1]$ are located in the surface plane. In this work, we present investgations on the surface magnetization of a Ni(110) single crystal, where we found a saturation magnetization in the topmost surface layer, which is not aligned along the easy axes $<$111$>$ as the deeper layers. 

The experiments were performed in an ultrahigh vacuum (UHV) chamber at a base pressure in the $10^{-11}$ mbar range, attached via two differential pumping stages to the beam lines of two different electrostatic ion accelerators with energies up to 30 kV, or, alternatively, up to 350 kV. The Ni(110)
single crystal ($12.8\times5.2\times4.0$) mm was prepared by cycles of grazing sputtering with 50 keV Ar$^+$ ions and subsequent annealing to 850 K for 30 minutes, until the surface was clean and flat as checked by low energy electron diffraction (LEED) using a SPA-LEED instrument (Omicron) and Auger electron spectroscopy (AES) using an electron gun (LEG32, VG-Scienta) and a CSA300 electron spectrometer (Omicron). The target temperature was controlled by a NiCr-Ni thermocouple attached close to the crystal. For the magnetic measurements, the Ni(110) crystal was placed in the gap of a softmagnetic FeCo yoke of a coil in order to remanently magnetize the crystal in a single-domain state of the saturation magnetization along the $[11\bar{1}]$ or $[\bar{1}\bar{1}1]$ easy axis in the (110) surface plane ("in-plane magnetization"). This procedure reproducibly yields a full remanent magnetization of the crystal as checked by magneto-optic Kerr effect (MOKE). For changing the azimuthal settings, the Ni(110) crystal and the FeCo yoke were mounted on a rotatable manipulator. External magnetic fields are compensated by three pairs of Helmholtz coils to a few $\mu$T.

After electron capture (EC) into excited states of fast He atoms during grazing scattering from Ni(110), the emitted polarized fluorescence light of the $1s3p~^3$P$\rightarrow\!1s2s~^3$S transition at $\lambda\!=\!388.9$~nm was detected through a quartz window by means of a quarter-wave retarder plate, a narrow bandwidth interference filter, a linear polarizer, and a cooled photomultiplier. The concepts and analysis of experiments on polarized light emission after electron capture are described in detail in Refs. \cite{Winter1,Winter2,Leuker1,Leuker2,Winter3,Winter4}. In brief, the spin polarization $P_S^\mathrm{EC}$ of captured electrons can be deduced from the circular polarization of the fluorescence light described by the Stokes parameter $S/I\!=\![I(\sigma^-)\!-\!I(\sigma^+)]/[I(\sigma^-)\!+\!I(\sigma^+)]$, where $I(\sigma^-)$ and $I(\sigma^+)$ are the intensities of light with negative and positive helicities, $\sigma^-$ and $\sigma^+$, respectively \cite{Guenther}. The spin polarization $P_S^\mathrm{EC}$ is obtained from measurements of the Stokes parameter $S/I(\uparrow)$ and $S/I(\downarrow)$ with reversed settings of the saturation magnetization, and is related to the long-range magnetic order of the topmost surface layer with a probing depth $\lambda_{\mathrm{EC}}\!\rightarrow\!0$~ML \cite{Pfandzelter1,Gruyters2,Busch1,Busch2}.

After energy separation by the CSA300 electron spectrometer, emitted secondary electrons induced by 2 keV electrons or by grazingly scattered 50 keV protons are imaged by an electrostatic lens onto a spin-polarized low-energy electron diffraction detector (SPLEED, Omicron) \cite{Kirschner1}. In this detector electrons are backscattered at a constant energy of 104.5~eV from a clean W(100) surface and the intensities of the $(2,0)$ and $(\bar{2},0)$ LEED spots are recorded with a pair of channeltrons. From the asymmetries of signals, caused by different cross-sections for left-right scattering, the "in-plane" component of the electron spin polarization can be deduced. In order to correct for instrumental asymmetries owing to different detector efficiencies, misalignment of the incident beam, etc., the electron spin polarization $P_S^\mathrm{SPLEED}$ is obtained from measurements under reversed magnetizations. For details concerning SPLEED measurements we refer to literature \cite{Kirschner1,Kirschner2,Siegmann,Busch3}.

The bulk magnetization of the crystal was observed by making use of MOKE in the longitudinal geometry. In order to record hysteresis loops, the change in the intensity of light from an electronically stabilized laser diode ($\lambda\!=\!635$ nm) that passes through an analyzing polarizer (set to an angle close to extinction) is monitored as the applied magnetic field is swept \cite{Liu}. The peak-to-peak intensity $\Delta I_\mathrm{MOKE}$, which is the difference in MOKE intensities, at positive and negative saturation magnetizations, is related to the amount of Kerr rotation and to the total magnetic moment \cite{Liu,Mentz}. The EC and MOKE measurements were performed for different pulse decay times in a range from 10~$\mu$s (designated as {\it fast current change}) up to 10~s ({\it slow current change}).

\begin{figure}[b]
\vspace{-0.4cm}
\includegraphics[width=7cm]{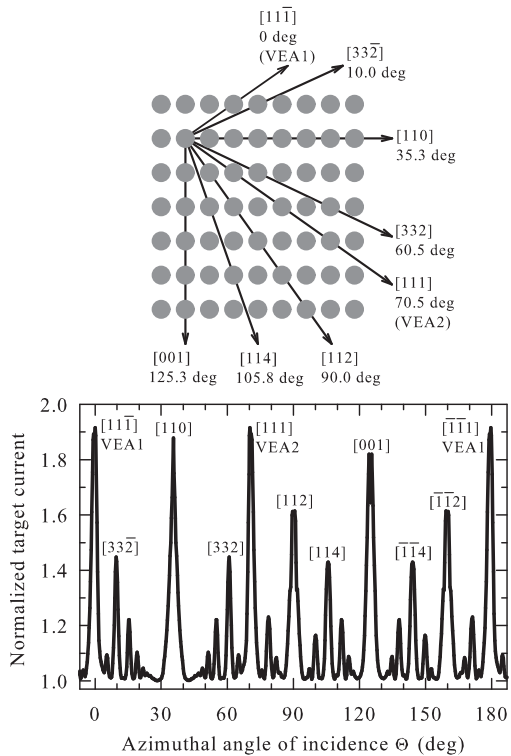}
\caption{Structure model of Ni(110) surface (top) and normalized target current versus azimuthal angle of incidence $\Theta$ for grazing scattering of 50 keV protons from Ni(110) under $\Phi_\mathrm{in}\!=\!0.9$ deg  at room temperature (bottom). Several low-index directions of fcc lattice are labeled by $[uvw]$. (VEA1: volume easy axis, which is collinear to magnetic field; VEA2: second volume easy axis.)}
\end{figure}

In order to identify specific directions in the Ni(110) surface, the target current as function of the azimuthal angle of incidence $\Theta$ was recorded. The measurement was performed at room temperature with grazingly scattered 50 keV protons under a polar angle of incidence $\Phi_\mathrm{in}\!\!=\!\!0.9$ deg with respect to the surface plane of the target. A typical result is shown in Fig. 1 (bottom). The target current is normalized to one for random azimuthal orientation. The incoming protons are steered by strings of surface atoms and may penetrate into subsurface layers, whenever they impinge along a low-index crystallographic direction in the surface plane {\it (axial surface channeling)} \cite{Winter5}. Compared to random azimuthal settings the resulting projectile trajectories lead to enhanced electron emission so that the number of emitted electrons as function of azimuthal angle $\Theta$ exhibits maxima at low-index directions $[uvw]$ of the fcc lattice \cite{Pfandzelter2}.

\begin{figure}[b]
\vspace{-0.6cm}
\includegraphics[width=8cm]{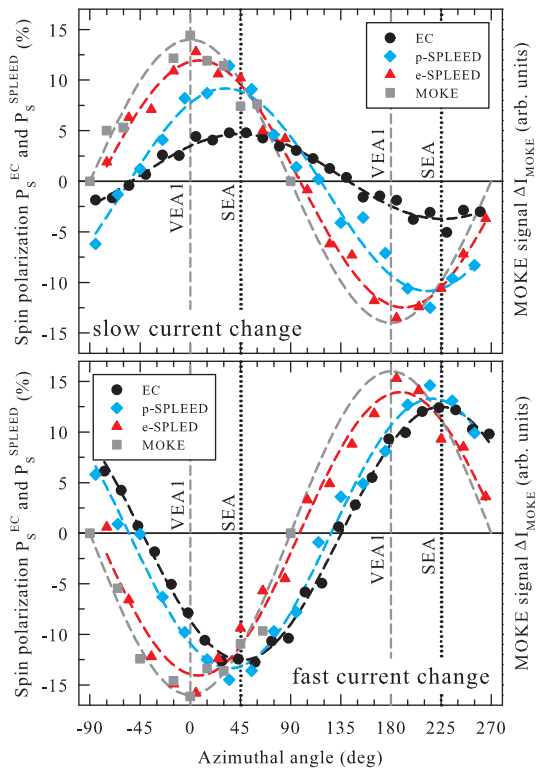}
\caption{(Color online) Spin polarization $P_S^\mathrm{EC}$ (left axis; black circles) deduced from electron capture for grazingly scattered 20 keV He$^+$ ions under $\Phi_\mathrm{in}\!=\!1.2$ deg, spin polarization $P_S^\mathrm{SPLEED}$ (left axis) of secondary electrons emitted by grazingly scattered 50 keV protons under $\Phi_\mathrm{in}\!=\!1.2$ deg (blue diamonds) and by 2 keV electrons under $\Phi_\mathrm{in}\!=\!35$ deg (red triangles), and MOKE signal $\Delta I_\mathrm{MOKE}$ (right axis; gray squares) versus azimuthal angle for slow (upper panel) and fast (lower panel) current change for remanent magnetization of Ni(110) measured at $T\!=\!300$ K. Dashed curves are sine functions as guide to eyes. (SEA: surface easy axis.)}
\end{figure}

\begin{figure}[b]
\vspace{-0.5cm}
\includegraphics[width=8cm]{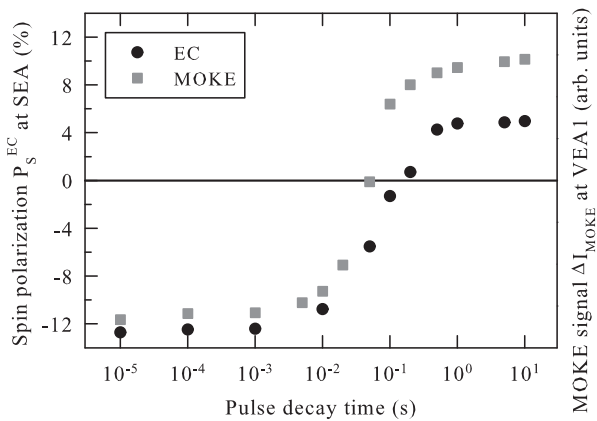}
\caption{Spin polarization $P_S^\mathrm{EC}$ (left axis, black circles) deduced from electron capture for 20 keV He$^+$ ions measured along SEA and MOKE signal $\Delta I_\mathrm{MOKE}$ (right axis, gray squares) measured along VEA1 versus pulse decay time of current for remanent magnetization of Ni(110) at $T\!=\!300$~K.}
\end{figure}

In Fig. 2 we show the spin polarization $P_S^\mathrm{EC}$ deduced from electron capture for grazingly scattered 20 keV He$^+$ ions under a polar angle of incidence $\Phi_\mathrm{in}\!=\!1.2$ deg, the spin polarization $P_S^\mathrm{SPLEED}$ of secondary electrons averaged over energies from 30 to 40 eV emitted by grazingly scattered 50 keV protons under $\Phi_\mathrm{in}\!=\!1.2$ deg and 2~keV electrons under $\Phi_\mathrm{in}\!=\!35$ deg, and the MOKE signal $\Delta I_\mathrm{MOKE}$ versus the azimuthal angle for "slow" as well as "fast" current change for remanent magnetization of Ni(110) measured at $T\!=\!300$ K (see below and Fig. 3). The data in both panels show the expected sine-dependence of $P_S^\mathrm{EC}$, $P_S^\mathrm{SPLEED}$, and $\Delta I_\mathrm{MOKE}$ on the azimuthal angle. The different sign between the data shown in the upper and lower panels results from the self-induction in the coil for fast (pulsing) change in current. For MOKE (gray squares) the maximum and minimum were found parallel or antiparallel with respect to the direction of magnetization along the (volume) easy axis VEA1 $[11\bar{1}]$ and $[\bar{1}\bar{1}1]$, which is collinear to the magnetic field of the FeCo yoke. However, the behavior of the spin polarization $P_S^\mathrm{EC}$ (black circles) is shifted by about 45 deg with respect to the MOKE data. Hence, the maximum and minimum of $P_S^\mathrm{EC}$ were not along VEA1 rather than along an axis, named here surface easy axis (SEA), oriented 10 deg next to the $<$110$>$ directions, which are the {\it medium axes} for Ni \cite{Kaya3}. This shift we found for slow as well as fast (pulsing) current change (cf. upper and lower panel of Fig. 2). The data of the proton-induced (blue diamonds) and electron-induced SPLEED measurements (red triangles) reveal the transition from the topmost surface layer to the bulk. Due to a probing depth $\lambda_{\mathrm{p-SPLEED}}\!\lesssim\!1$~ML \cite{Potthoff1} for non-penetrating, grazingly scattered protons the behavior of the spin polarization $P_S^\mathrm{p-SPLEED}$ is shifted by about 30 deg with respect to the MOKE data, whereas  $P_S^\mathrm{e-SPLEED}$ (with $\lambda_{\mathrm{e-SPLEED}}\!=\!5-7$~ML, \cite{Potthoff1}) is shifted by about 7 deg only with respect to the MOKE data. The same anomalous orientation of the saturation magnetization of the topmost surface layer was observed for a circularly shaped Ni(110) single crystal (diameter 10 mm, 5 mm thick). 

In order to investigate the transition between the regimes of slow and fast (pulsing) current change, we performed EC and MOKE measurements for different pulse decay times in a range from 10~$\mu$s up to 10~s. The pulse rise times were equal to the pulse decay times and the pulse width was 500 ms in all cases. The results for $P_S^\mathrm{EC}$ (black circles) measured along the surface easy axis SEA and for MOKE signal $\Delta I_\mathrm{MOKE}$ (gray squares) measured along the volume easy axis VEA1 as function of the pulse decay time of the current for remanent magnetization of Ni(110) at $T\!=\!300$~K are displayed in Fig. 3. Both data sets show a comparable time dependence and a saturation for pulse decay times below 10~ms and above 1~s. For a pulse decay time of about 0.1~s $P_S^\mathrm{EC}$ and $\Delta I_\mathrm{MOKE}$ show a crossover. Hence, the current change has no effect on the observed anomalous magnetic anisotropy.

\begin{figure}[b]
\vspace{-0.7cm}
\includegraphics[width=8cm]{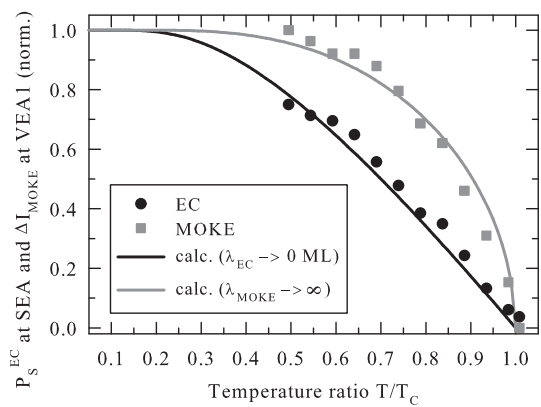}
\caption{Normalized Spin polarization $P_S^\mathrm{EC}$ (black circles) deduced from electron capture for 20 keV He$^+$ ions measured along SEA and normalized MOKE signal $\Delta I_\mathrm{MOKE}$ (gray squares) measured along VEA1 for fast current change for remanent magnetization of Ni(110) versus temperature ratio $T/T_C$ (with $T_C\!=\!627$~K, \cite{Heller}). Curves represent mean-field calculations with different probing depths $\lambda$ as indicated.}
\end{figure}

A characteristic feature of ferromagnetic materials is the decay of spontaneous magnetization with increasing temperature $T$ until the paramagnetic state is reached at the Curie temperature $T_C$ (for Ni: $T_C\!=\!627$~K, \cite{Heller}). In the case of a band ferromagnet, such as Fe or Ni as prototypes, the temperature dependence of the spontaneous magnetization can be explained by the electronic structure. At the surface of a band ferromagnet, the magnetization may be different due to the reduced translational symmetry. Within the framework of classical spin models, the lowered surface coordination implies that the magnetization of the topmost surface layer is substantially reduced as compared to the bulk \cite{Potthoff3,Herrmann}. However, significant deviations from the temperature dependence of the bulk magnetization are confined to the upper most surface layers. Nevertheless, the detailed temperature dependence of the spontaneous magnetization for a band-ferromagnet surface must still be considered as an open issue. 

Pfandzelter and Potthoff investigated the temperature- and layer-dependent magnetization of a Fe(100) single crystal via EC, proton- and electron-induced SPLEED, and MOKE measurements and found good accordance of the experimental data with mean-field calculations \cite{Potthoff1}. We also performed EC and MOKE measurements on Ni(110) in a temperature range between 300 K and $T_C\!=\!627$ K. In Fig. 4 we show the normalized spin polarization $P_S^\mathrm{EC}$ measured along SEA and MOKE signal $\Delta I_\mathrm{MOKE}$ measured along VEA1 for fast current change for remanent magnetization of the Ni(110) surface versus the temperature ratio $T/T_C$. Irrespective of the observed anomalous orientation of the magnetization of the topmost surface layer, we found for Ni(110) good accordance between experimental data and results of mean-field calculations for extreme probing depths $\lambda_{\mathrm{EC}}\!\rightarrow\!0$~ML and $\lambda_{\mathrm{MOKE}}\!\rightarrow\!\infty$, as shown in Fig. 4.

\begin{figure}[t]
\vspace{-0.4cm}
\includegraphics[width=8cm]{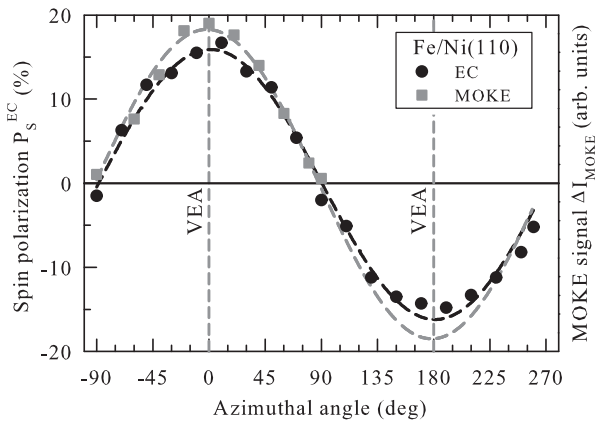}
\caption{Spin polarization $P_S^\mathrm{EC}$ (left axis; black circles) deduced from electron capture for 20 keV He$^+$ ions and MOKE signal $\Delta I_\mathrm{MOKE}$ (right axis; gray squares) versus azimuthal angle for fast current change for remanent magnetization of a 50 ML Fe film on Ni(110) measured at $T\!=\!300$~K. Dashed curves are sine functions as guide to eyes.}
\end{figure}

The anomalous orientation of the magnetization of the topmost surface layer of Ni(110) as found in this work was not observed in a former study on a Fe(110) single crystal reported by Leuker {\it et al.} \cite{Leuker2}. For Fe(110) the maximum and minimum of the spin polarization $P_S^\mathrm{EC}$ is parallel and antiparallel, respectively, with respect to the direction of magnetization along the easy axes $<$100$>$. Furthermore, it was deduced from the observed sine-dependence that $P_S^\mathrm{EC}$ is not affected by axial channeling effects. As consistency check, we have grown a 50~ML film Fe on Ni(110) at room temperature and performed EC and MOKE measurements. The results are shown in Fig. 5. Contrary to Ni(110), the azimuthal angular dependence of $P_S^\mathrm{EC}$ and $\Delta I_\mathrm{MOKE}$ for 50~ML Fe on Ni(110) is the same.

In summary, we have investigated the orientation of magnetization for a Ni(110) surface using techniques with different probing depths. By making use of electron capture measurements with $\lambda_{\mathrm{EC}}\!\rightarrow\!0$~ML we found that the magnetization of the topmost surface layer is not aligned along the (volume) easy axes $<$111$>$. To our knowledge, this is the first observation of this feature. In accord with previous investigations on a Fe(110) single crystal, we found for a 50~ML film Fe on Ni(110), that maximum and minimum of the spin polarization $P_S^\mathrm{EC}$ are oriented along the easy axes $<$100$>$ of Fe. A theoretical understanding of the observed anomalous magnetic anisotropy on Ni(110) is beyond the scope of the present work, but should be accessible to state-of-the-art {\it ab initio} calculations based on density-functional theory.
\acknowledgements
We thank K. Maass and G. Lindenberg for their assistance in the preparation of the experiments. We are grateful to W. Nolting (Berlin) for helpful discussions.

\end{document}